\documentclass[preprint,aps,nofootinbib]{revtex4}
\usepackage{amssymb}
\usepackage{graphics}
\begin{document}

\newcommand{\be}{\begin{equation}}
\newcommand{\ee}{\end{equation}}
\def\bq{\begin{eqnarray}}
\def\eq{\end{eqnarray}}


\title{\bf Gravity of $R^{\mu\nu}=0$: A New Paradigm in GR\footnote{Awarded {\sc `Honorable Mention'} of the year 2012 by the Gravity Research Foundation}}
\author{Ram Gopal Vishwakarma}

 \address{Unidad Acad$\acute{e}$mica de Matem$\acute{a}$ticas\\
 Universidad Aut$\acute{o}$noma de Zacatecas\\
 C.P. 98068, Zacatecas, ZAC,
 Mexico\\
Email: vishwa@matematicas.reduaz.mx}

\begin{abstract}
Theory of general relativity (GR) has been scrutinized by  experts for almost a century and describes accurately all gravitational phenomena ranging from the solar system
to the universe. However, this success is achieved provided one admits  three completely independent new components in the energy-stress tensor $T^{\mu\nu}$$-$inflaton, dark matter and dark energy, which though do not have any non-gravitational or laboratory evidence and have remained generally speculative. Moreover, the dark energy poses a serious confrontation between fundamental physics and cosmology.

The present situation reminds us of Einstein's `biggest blunder' when he forced his theory to predict a static universe, perhaps guided by his religious conviction that the universe must be eternal and unchanging. It seems that we are making a similar blunder by forcing  $T^{\mu\nu}$ into the field equations while the observations indicate that it is not needed.
We seem to have a deep-rooted conviction that the spacetime will remain empty unless we fill it by the  energy-stress tensor. However, we have been ignoring numerous evidences earnestly indicating otherwise. 

 From a critical analysis of the present situation, we develop an entirely new insight about the source of curvature in equations $R^{\mu\nu}=0$ which, though may appear orthogonal to the usual understanding, is in striking agreement with all known phenomena in GR. Moreover, it answers some hitherto unexplained puzzles and circumvents some long-standing problems of the standard paradigm.

\noindent
{\bf Key words:} General Relativity and Gravitation - theory - fundamental problems and general formalism - cosmological observations. 

\pacs{04.20.Cv, 04.20.-q, 95.30.Sf, 98.80.Jk}

\end{abstract}
\maketitle

\section{Introduction}

Einstein's theory of general relativity (GR) ranks as one of the crowning intellectual achievements of the twentieth century.
Although GR is not the only relativistic theory of gravitation, it is the simplest theory that has survived the tests of nearly a century of observational confirmation ranging from  the solar system to the largest scales, the universe itself.
 However, this success is achieved provided we admit  three
completely independent new components in the energy-stress tensor $T^{\mu\nu}$ $-$  inflaton, dark matter and dark energy, which are believed to play major roles in the dynamics of the universe during their turns.  However, there is, until now, no non-gravitational or laboratory
evidence for any of these dark sectors. Admittedly,  the conditions of the early universe cannot be brought back to directly observe inflaton, but neither has been observed any relic thereof. Neither the dark matter has any confirmed observational evidences, though the efforts to directly observe it are still going on. The dark energy is the most enigmatic  among all the dark sectors which poses a serious
confrontation between fundamental physics and cosmology. The most favoured candidate of the dark energy - the cosmological constant - is plagued with the long-standing `cosmological constant problem'.

The origin of the cosmological constant problem lies in a conflict between the energy-stress tensor $T^{\mu\nu}$ and the quantum field theory (QFT):
The vacuum energy, according to the QFT, results from the quantum vacuum fluctuations which provide an energy contribution of the order of the Planck mass. In GR, the vacuum energy is represented by $T^{\mu\nu}$ with a particular equation of state $p_{\rm vac}=-\rho_{\rm vac} c^2$ (i.e., through the cosmological constant). Einstein's theory (through the Friedman equation) then provides an estimate of the vacuum energy of the order of $H_0^2$, where $H_0$ is the present value of the Hubble parameter. This is, however, smaller than the QFT-value by a factor of $\approx 10^{-120}$!
(This discrepancy has been called `the worst theoretical prediction in the history of physics!')

A huge value of the vacuum energy is also predicted by inflation which requires it to expand the early universe by a factor of $10^{78}$ in just $10^{-36}$ seconds, leaving a “nearly” flat spacetime. However, a flat spacetime, which is a notion of special relativity (SR), is not compatible with the real universe in the presence of matter in the existing framework of GR. The origin of this problem is again the energy-stress tensor  $T^{\mu\nu}$.
As the derivation of $T^{\mu\nu}$ assumes its validity in the absence of gravitation in SR\footnote{Let us recall that the general expression of the tensor $T^{\mu\nu}$ is obtained by first deriving it in SR. The bridge between the ideal case of SR and the actual case in the presence of gravity, is provided by an inertial observer, which exists admittedly at all points of spacetime (by courtesy of the principle of equivalence). The fluid is defined in a small neighbourhood of the observer. Then, the expression of the tensor, in the presence of gravity, is imported from SR through a coordinate transformation.}, this goes contradictory to the very notion of $T^{\mu\nu}$ being the source of curvature.

Thus, besides its incompatibility with the other three forces of nature, GR seems to face the following three problems.

\begin{enumerate}
\item
Admittance of the speculative dark sectors in $T^{\mu\nu}$ without any direct experimental support.

\item
The cosmological constant problem, which appears through the presence of $T^{\mu\nu}$ in Einstein's equations.

\item
No scope for a flat spacetime in the presence of matter.

\end{enumerate}

Surprisingly all these problems are somehow related with the energy-stress $T^{\mu\nu}$, which has become an integral part of the modern theories of gravitation, including the candidate theories of quantum gravity.
The present situation reminds us of Einstein's `biggest blunder' when he forced his theory to predict a static universe, perhaps guided by his religious conviction that the universe should be eternal and unchanging. It seems that a similar blunder is being made by forcing  $T^{\mu\nu}$ into the field equations. We seem to have a deep-rooted conviction that the spacetime will remain empty unless we fill it by the  energy-stress tensor, despite numerous evidences earnestly indicating otherwise and the observations hinting that $T^{\mu\nu}$ is not needed.
Interestingly, all the problems mentioned above, including many more, can be avoided if one drops $T^{\mu\nu}$ from  Einstein's field equations resulting in the so-called `vacuum' field equations 
\be
R^{\mu\nu}=0.\label{eq:RicciEq}
\ee
As we shall see later, this equation is strongly supported, not only by the observations of the local universe (through various experiments performed to test GR), but also by the cosmological observations. It is believed that equations (\ref{eq:RicciEq}) cannot represent the actual universe as they represent an empty spacetime. Let us see if this is so.

\section{A New Insight About the Source of Curvature}

In an empty space in the absence of any source of curvature, one should expect a flat spacetime as a unique solution of field equations (\ref{eq:RicciEq}). However,  it has already been 
realized that this is not true, and in a space with dimensions four 
or more,  equations (\ref{eq:RicciEq}) can have curvature. A brilliant example of this case is the Schwarzschild solution which represents the spacetime structure outside an isotropic mass $m$. In the Schwarzschild coordinates\footnote{In Schwarzschild coordinates, $t$ is the time coordinate (measured by a stationary clock located infinitely far from the origin $r=0$) and $r,\theta, \phi$ are the spherical polar coordinates. The radial coordinate $r$ is measured as the circumference, divided by $2\pi$, of a sphere centered around $r=0$. As the space may not be Euclidean, we cannot claim that $r$ is the `radial distance' from the origin. Rather, $r$ is simply an arbitrary radial coordinate scaled to give the usual Euclidean circumference.}, the solution is given by
\be
ds^2=\left(1-\frac{2Gm}{c^2 r}\right)c^2 dt^2-\frac{dr^2}{1-2Gm/(c^2 r)}-r^2d\theta^2-r^2\sin^2\theta ~d\phi^2.\label{eq:Sch}
\ee
 The source of curvature in (\ref{eq:Sch}) is attributed to the mass $m$ sitting at $r=0$, which appears in the guise of a singularity in GR.  However, it should be noted that the metric
(\ref{eq:Sch}) represents space exterior to the central mass at $r=0$ and {\it not} the 
point $r=0$ itself, where the metric breaks down. So, how can a mass situated 
at the point $r=0$ (which is not even represented by the metric) curve the space of (\ref{eq:Sch}) at the points for which $r>0$? Obviously, the agent responsible for the curvature in  (\ref{eq:Sch}) at the points for $r>0$, must be the gravitational energy, which can definitely exist in an empty space. It is then clear that the spacetime of $R^{\mu\nu}=0$, as represented by one of its solutions, viz. (\ref{eq:Sch}), {\it does have energy, and is not empty}.

The important point to note is that equations $R^{\mu\nu}=0$ reveal the gravitational energy without containing any formulation thereof  (neither $T^{\mu\nu}$ contains the gravitational energy, as a proper energy-stress tensor of the gravitational field does not exist). This then implies that the gravitational energy already exists there implicitly in the geometry, through the non-linearity of the field equations (\ref{eq:RicciEq}), and no additional incorporation thereof is needed. This fits very well in
 the story of the failure to discover the energy-stress tensor of the gravitational field\footnote{It may be mentioned that despite a century-long dedicated efforts of many luminaries, the attempts to discover a unanimous formulation of the energy-stress tensor of the
gravitational field have failed. Primarily, because of the the non-tensorial character of the energy-stress pseudo-tensors of the gravitational field and the lack of a unique agreed-upon formula for it. Secondly, because of the inherent difficulty in the localization of the gravitational energy.}. 
A proper energy-stress tensor of the gravitational field does not exist simply because it
is not needed in the geometric framework of GR, it already exists there inherently in 
the geometry.

One can argue that  equations $R^{\mu\nu}=0$ can represent only empty space (except for having the gravitational energy), outside massive objects. However, this conviction does not seem correct and  equations $R^{\mu\nu}=0$ can also support curved
cosmological solutions valid through the whole span of spacetime.  The Kasner solution may be considered as an example:
\be
ds^2=c^2 dt^2- t^{2p_1}dx^2- t^{2p_2}dy^2- t^{2p_3}dz^2,\label{eq:kasner}
\ee
where the constants $p_1$, $p_2$ and $p_3$ satisfy
\[
p_1+p_2+p_3=1, ~~p_1p_2+p_2p_3+p_3p_1=0.
\]
The usual interpretation of solution (\ref{eq:kasner}) is provided in terms of an empty homogeneous universe in which the space is expanding/contracting anisotropically at different rates in different directions (for example, the space is expanding in two directions and contracting in the third).
Hence, in the absence of matter, the only other possible source of curvature in this solution, can be a singularity. The solution  (\ref{eq:kasner}) does contain a singularity at $t=0$, but {\it not at any other time}. However, the solution is curved at all times! A past singularity, which does not exist now, fueling the gravitational energy  now without any other source, does not seem compatible with our understanding of the gravitational energy.

The usual source of curvature in the Kasner metric (\ref{eq:kasner}) is regarded a net non-zero momentum resulting from the anisotropic expansion/contraction of the homogeneous space, since the space expands and contracts at different rates in different directions in it. However, how can nothing expand/contract? It does not make sense to imagine of momentum resulting from the expanding/contracting {\it empty} space.   Hence, the Kasner solution has remained an unexplained puzzle. 

An important point regarding the Kanser solution, which has not been paid attention to, is that unlike the Schwarzschild solution, it represents a cosmological solution, which is not expected to have any `outside'. Since the ultimate source of the gravitational field is matter, the (homogeneously distributed) matter source, present at the time of singularity in the Kanser metric, must be present at all other times as well, as it {\it must not} have been destroyed mysteriously! 
This simply means that the Kasner solution represents a homogeneous distribution of matter expanding and contracting anisotropically!

This implies that, like the energy and momentum of the gravitational field, those of the matter fields are also included in equations $R^{\mu\nu}=0$ inherently (without including any additional formulation thereof) whose effects are revealed through the geometry. 
The futility of the energy-stress tensor is also corroborated by the Kerr solution  wherein the angular momentum also contributes to its curvature:
\[
ds^2=\left(1-\frac{r_{\rm S}r}{\rho^2}\right)c^2 dt^2-\frac{\rho^2}{\Delta}dr^2-\rho^2d\theta^2~~~~~~~~~~~~~~~~~~~~~~~~~~~~~~~~~~~~~~~~~~~~~~~~
\]
\be
 -\left(r^2+\alpha^2+\frac{r_{\rm S}r\alpha^2}{\rho^2}\sin^2\theta\right)\sin^2\theta ~d\phi^2+\frac{2r_{\rm S}r\alpha}{\rho^2}\sin^2\theta ~d\phi ~cdt,\label{eq:kerr}
\ee
which describes the spacetime surrounding a spherical mass $m$  spinning with angular momentum per unit mass = $\alpha$ (so that its total angular momentum $=mc\alpha$). Here $\rho^2=r^2+\alpha^2 \cos^2\theta$, $\Delta=r^2-r_{\rm S}r+\alpha^2$ and $r_{\rm S}=2Gm/c^2$ is the Schwarzschild radius.
 It may be mentioned that there is no place for the angular momentum in $T^{\mu\nu}$ in the framework of Einstein's theory\footnote{which needs to be extended to non-Riemannian curved spacetime with torsion (as in the Einstein-Cartan theory) to support asymmetric Ricci and metric tensors, so that an asymmetric energy-stress tensor of spin can appear on the right hand side of the equations.}. But, since angular momentum does provide a source of curvature/gravitation, as is evidenced by (\ref{eq:kerr}), this provides another reason why  $T^{\mu\nu}$ should not appear in the field equations.

 Einstein believed that \cite{einstein}

\begin{quote}
{\it ``On the basis of the general theory of relativity, space as opposed to `what fills space', which is dependent on the coordinates, has no separate existence.  ... The functions $g_{\mu\nu}$ describe not only the field, but at the same time also the topological and metrical structural properties of the manifold.  ... Spacetime does not claim existence on its own, but only as a structural quality of the field''.} 
\end{quote}

\noindent
Thus the mere consideration of a spacetime structure (conditioned by the equations $R^{\mu\nu}=0$) must be equivalent to considering the accompanying fields (material and gravitational) as well, and there should be no need to add any extra formulation thereof to the field equations. Hence, the structure of the geometry is determined by the net contribution from the material and the gravitational fields (of the chosen matter distribution).
This new discovery might appear orthogonal to the usual understanding of Einstein's theory. Nevertheless, it is in perfect agreement with all gravitational phenomena encountered in GR.  Moreover, it also provides natural explanations to some unexplained puzzles of GR, as we shall see in the following. 

The above-gained insight about the implicit presence of material and gravitational fields in  equations $R^{\mu\nu}=0$ makes a powerful prediction: As the source of curvature in the Kasner solution (\ref{eq:kasner}) is a net non-zero momentum resulting from the anisotropic expansion/contraction of the homogeneous material distribution, one should expect a flat spacetime as a solution of equations $R^{\mu\nu}=0$ for a homogeneous matter distribution expanding or contracting isotropically. 
This prediction is perfectly realized in the following cosmological solution, which provides a concrete evidence for the correctness of our new discovery about the source of curvature in GR.
 Obviously, the symmetries of a homogeneous matter distribution expanding or contracting isotropically require the metric
to be the Robertson-Walker one given by
\be
ds^2=c^2 dt^2-S^2\left(\frac{dr^2}{1-kr^2}+r^2d\theta^2+r^2\sin^2\theta ~d\phi^2\right),\label{eq:RW}
\ee
where $S(t)$ is the scale factor of the universe. For this metric, equations (\ref{eq:RicciEq})  yield
\be
R^0_{~0}=\frac{3}{c^2}\frac{\ddot{S}}{S}=0,
\ee
\be
R^1_{~1}=R^2_{~2}=R^3_{~3}=\frac{1}{c^2}\left(\frac{\ddot{S}}{S}+2\frac{\dot{S}^2}{S^2}+2kc^2\frac{1}{S^2}\right)=0,
\ee
which uniquely determine
\be
S=ct ~~\text{with} ~~ k=-1,\label{eq:scale}
\ee
so that the final solution reduces to
\be
ds^2=c^2 dt^2-c^2t^2\left(\frac{dr^2}{1+r^2}+r^2d\theta^2+r^2\sin^2\theta ~d\phi^2\right).\label{eq:milne}
\ee
It may be noted that by the use of the transformations $\bar{t}=t\sqrt{1+r^2}$, $\bar{r}=ctr$, the solution (\ref{eq:milne}) can be brought to manifestly Minkowskian form in the coordinates $\bar{t},\bar{r}, \theta, \phi$ \cite{narlikar}.

It would be natural to ask why solution (\ref{eq:milne}) is flat while solutions  (\ref{eq:Sch})$-$(\ref{eq:kerr}), of the same field equations $R^{\mu\nu}=0$, are curved. 
One may argue that the reason one gets the Minkowskian solution (\ref{eq:milne}) is simply because $T^{\mu\nu}$ is vanishing in  (\ref{eq:RicciEq}). However, if this is so, why do we get curved solutions (\ref{eq:Sch})$-$(\ref{eq:kerr}) from the same equations (\ref{eq:RicciEq}) without imposing any extra condition? If equations  $R^{\mu\nu}=0$ represent a non-empty space in solutions (\ref{eq:Sch})$-$(\ref{eq:kerr}), they must do so in solution (\ref{eq:milne}) also. 
One may further argue that solutions (\ref{eq:Sch})$-$(\ref{eq:kerr}) have singularities and so they are curved, while the solution (\ref{eq:milne}) does not have a singularity, and so it is not curved. But, what is there to stop the singularity to occur in (\ref{eq:milne})? The only difference between the considerations leading to solutions (\ref{eq:Sch})$-$(\ref{eq:kerr}) and  (\ref{eq:milne}) is that we have assumed different types of symmetries in their spacetime structures. While the metric (\ref{eq:milne}) is homogeneous and isotropic, the metrics   (\ref{eq:Sch})$-$(\ref{eq:kerr}) are either inhomogeneous or/and anisotropic. However, how a relaxation in the homogeneity and/or isotropy can result in a singularity, cannot be explained by  the conventional wisdom.
 For example, solutions (\ref{eq:kasner}) and (\ref{eq:milne})  represent 
similar spacetime structures with the only difference that while the homogeneous space
in (\ref{eq:kasner})  is expanding and contracting in different directions at different rates, the same
space is expanding or contracting isotropically in (\ref{eq:milne}). How does this difference account for
their curved and  flat states and controls the appearance of the singularity? These questions cannot be answered satisfactorily in the framework of the conventional wisdom.

A convincing  and natural explanation emerges from the above-made discovery that the sources of curvature (the material and the gravitational fields) are inherently present in the geometry of  equations (\ref{eq:RicciEq}). According to this, solution (\ref{eq:milne}) represents a spacetime structure resulting from the homogeneously distributed matter throughout the space at all times, expanding or contracting isotropically. Then, why is it not curved? Simply because the positive energy of the matter field is exactly balanced, point by point,  by the negative energy of the resulting gravitational field  (contrary to the case of the Schwarzschild metric where there is only the gravitational energy and no matter at the points represented by the metric), providing a net vanishing energy. Neither there is any net non-zero momentum contribution from the isotropic expansion or contraction of the material system (contrary to the case of the Kasner metric).  Hence, in the absence of any net non-zero energy, momentum or angular momentum, the spacetime of (\ref{eq:milne}) must not have any curvature.

It thus appears that it is the  symmetry of the chosen spacetime structure, which determines whether the solution of $R^{\mu\nu}=0$ will be curved (may possess a singularity) or flat (may not possess a singularity).
 This fact is also reflected in the appearance of different kinds of singularities in accordance with the chosen symmetries in solutions (\ref{eq:Sch})$-$(\ref{eq:kerr}): while the Schwarzschild solution (describing the spacetime structure exterior to a point mass) has a point singularity, the Kerr solution (describing the spacetime structure exterior to a rotating mass) has a ring singularity, the Kasner solution (in which the $t$ = constant hyper-surfaces are expanding and contracting in different directions at different rates) presents a singularity of the oscillating kind at $t=0$. 

The discovery, of the net vanishing  energy-momentum-angular momentum in a homogeneous distribution of matter expanding or contracting isotropically, appears consistent with several investigations and results which indicate that the total energy of the universe is zero. Thus, here we get a flat spacetime solution in the presence of matter, which originates dynamically from the field equations, and is not assumed a priori (or put by hand) as in SR.
It is thus clear that equations $R^{\mu\nu}=0$ constitute consistent field equations of gravitation even in the presence of matter. Further, a solution of $R^{\mu\nu}=0$  is curved when the conditions of homogeneity and/or isotropy are relaxed, otherwise the curvature is lost if the solution is homogeneous and isotropic.

\section{Support from the Cosmological Observations to $R^{\mu\nu}=0$}

As the last words on any physical theory is to be spoken by the observations/experiments, let us see how equations  $R^{\mu\nu}=0$  fair against the observations. The validity of the equations has already been well-established by the local observations through the classical tests of GR. Most of these tests are treated with the solutions of $R^{\mu\nu}=0$ obtained under the simplifying assumptions of isotropy\footnote{except for the recently made Gravity Probe B experiment which involves anisotropic effects owing to the rotation of the earth and is treated with solution (\ref{eq:kerr}).} and time-independence, which describe, to a good approximation, the gravitational field around the sun. Let us
study the compatibility of $R^{\mu\nu}=0$ with the cosmological observations. For this purpose let us consider its homogeneous-isotropic solution (\ref{eq:milne}), as would be expected from the observations at a sufficiently large scale. 
It may be mentioned that there is a genuine controversy surrounding the interpretation of redshift in cosmology, leading to various alternative interpretations thereof. However, in order to compare our results with those obtained in the standard cosmology, we  consider the standard interpretation of redshift given in terms of the cosmological expansion.

\bigskip
\noindent
{\bf Observations of Supernovae Ia}

\noindent
In order to study the compatibility of equation (\ref{eq:milne}) with the cosmological observations, let us first consider the observations of supernovae of type Ia (SNeIa). It can be checked that solution (\ref{eq:scale}) is efficient enough to define uniquely, without requiring any inputs from the matter fields, different distance measures, for example, the
luminosity distance $d_{\rm L}$ and the angular diameter distance $d_{\rm A}$, in this theory.

 It is already known that the model based on equation (\ref{eq:milne}), albeit non-accelerating (neither decelerating), is consistent with the observations of SNeIa {\it without requiring any dark energy}.
As early as in 1998, the Supernova Cosmology Project team noticed from the analysis of their first-generation of the SNeIa data that `the performance of the empty model ($\Omega_{\rm m}=0=\Omega_\Lambda$) is practically identical to that of the best-fit unconstrained cosmology with a positive $\Lambda$' \cite{perlmutter}.
 Let us consider a newer dataset, for example, the  `new gold sample' of 182 SNeIa \cite{riess}\footnote{Although various newer SNeIa datasets are available, however, the way they are analyzed has left little scope for testing a theoretical model with them. This issue has been addressed in \cite{critique}.}, which is a reliable set of SNeIa with reduced calibration errors arising from the systematics. 
The present model provides an excellent fit to the data with a value of $\chi^2$ per degrees of freedom (DoF) $=174.29/181=0.96$ and a probability of the goodness of fit $Q=63\%$. Obviously the standard $\Lambda$CDM model has even a better fit as it has more free parameters: $\chi^2$/DoF $  =158.75/180=0.88$ and $Q=87\%$ obtained for the values $\Omega_{\rm m}=1-\Omega_\Lambda=0.34\pm0.04$.

\bigskip
\noindent
{\bf Observations of High-Redshift Radio Sources}

\noindent
Let us now consider the data on the angular size and redshift of radio sources compiled by Jackson and Dodgson \cite{radio},
which has 256  sources with their redshift in the range 0.5 - 3.8. These sources are ultra-compact radio objects
 of angular sizes of the order of a few milliarcseconds, deeply embedded
in the galactic nuclei and have very short lifetime compared with the age of the universe. Thus
they are expected to be free from evolutionary effects and hence may be treated as standard
rods, at least in the statistical sense. These sources are distributed into 16 redshift bins, each
bin containing 16 sources. This compilation has recently been used by many authors to test
different cosmological models \cite{radios-used}.

We find that the present model has a satisfactory fit to the data with $\chi^2$/DoF $= 20.78/15=1.39$ and $Q = 14\%$.
In order to compare, we find that the best-fitting $\Lambda$CDM model has a slightly better fit: $\chi^2$/DoF $= 16.03/14=1.15$ and $Q = 31\%$ obtained for the values $\Omega_{\rm m}=1-\Omega_\Lambda=0.21\pm0.08$.

\bigskip
\noindent
{\bf Observations of $H_0$ and $t_{\rm GC}$}

\noindent
The age of the universe $t_0$, in the big bang-like theories,  is the time elapsed since the big bang. It depends on the expansion dynamics of the model and is given by
\be
t_0=\int_0^\infty \frac{H^{-1}(z)}{(1+z)}dz.\label{eq:age}
\ee
Hence, the Hubble parameter controls the age of the universe, which in turn depends on the free parameters of the model.
For example, in the standard cosmology, $H(z) =H_0\{\Omega_{\rm m}(1+z)^3+\Omega_\Lambda+(1-\Omega_{\rm m}-\Omega_\Lambda)(1+z)^2\}^{1/2}$. Thus, by using the observed value of  $H_0$, one can calculate the age of the universe predicted by a particular theory.
As the universe is expected to be at least as old as the oldest objects in it, a lower limit is put on $t_0$. This is done through $t_{\rm GC}$, the age of the globular clusters in the Milky Way which are among the oldest objects we so far know. 
 Thus the measurements of $H_0$ and $t_{\rm GC}$ provide a powerful tool to test the underlying theory. 

For instance, by using the current measurements of $H_0 = 71\pm 6$ km s$^{-1}$ 
Mpc$^{-1}$ from the Hubble Space Telescope Key Project \cite{HST},  equation (\ref{eq:age}) gives $t_0$ for the Einstein-deSitter model ($\Omega_{\rm m}=1$, $\Lambda=0$) as $9.18$ Gyr. This cannot be reconciled with the age of the oldest globular cluster estimated to be  $t_{\rm GC}=12.5 \pm 1.2$ Gyr \cite{Gnedin} and the age of the Milky Way as $12.5 \pm 3$ Gyr coming from the latest uranium decay estimates \cite{Cayrel}.
However, for the concordance $\Lambda$CDM model with $\Omega_{\rm m}=1-\Omega_\Lambda=0.27$ (as estimated by the WMAP project \cite{wmap}), equation (\ref{eq:age}) gives a satisfactory age of the universe $t_0=13.67$ Gyr which is well above the age of the globular clusters.  The age of the universe in the present model is given by $t_0=H_0^{-1}$, as can be checked from (\ref{eq:scale}). For the above-mentioned value of $H_0$, this gives $t_0=13.77$ Gyr which is even higher than the concordance model value.

\bigskip
\noindent
{\bf Observations of CMB and BAO}

\noindent
Any proposed model of the universe is expected to provide a consistent theory of the structure formation and hence it is expected to explain the observations of the cosmic microwave background (CMB) radiation and the baryon acoustic oscillations (BAO). Though a detailed study of this subject would be out of scope of the essay, it may be mentioned that taking at the face values, the only unanimous prediction of the CMB observations is a flat geometry (of the $t=$ constant hyper-surface) \cite{wmap, cmb}. In this connection, it should be noted that equation (\ref{eq:milne}) can be brought to the Minkowskian form by the use of the suitable transformations, as has been mentioned earlier. In the new form, the metric obviously has a flat spatial geometry.

Additionally, as we have shown, the universe is not empty in the present model, though the matter fields do not play a direct role, hence providing full leverage on the parameters $\Omega_{\rm m}$, $\Omega_{\rm b}$, etc., to fit the observations of CMB and BAO  which point out that $\Omega_{\rm m}\approx 0.3$ \cite{CMB&BAO}.

\section{Conclusion}

In whatever manner we interpret the curvature appearing in a variety of solutions of $R^{\mu\nu}=0$, the important point is that this curvature appears without incorporating any formulation of the source (gravitational or material) into equations $R^{\mu\nu}=0$. Further, if the source of curvature, according to the metric theories of gravitation, is necessarily energy-matter, then the existence of non-zero curvature in these solutions guarantees the presence of energy-matter, implying that equations $R^{\mu\nu}=0$ {\it do not represent an empty spacetime.} 
Moreover, if this is true in one situation, this must be true in all the gravitational situations. This simply means that  the sources (gravitational fields as well as the matter fields) are inherently present in the equations (perhaps through their non-linearity), whose effects are revealed through the geometry. 

The fact that the sources of curvature are implicitly present in equations $R^{\mu\nu}=0$ and must not be added again (through the energy-stress tensor), is vindicated by the failure to obtain a proper energy-stress tensor of the gravitational field. It is further supported by a number of paradoxes and inconsistencies discovered recently in the relativistic formulation of matter given by the energy-stress tensor $T^{\mu\nu}$ \cite{vishwaApSS2} implying that, akin to the case of the gravitational field, a flawless proper energy-stress tensor of the matter fields neither exists.

This, in fact, leaves equation $R^{\mu\nu}=0$ as the only possibility for a consistent field equation of gravitation.
It is generally believed that a consistent field equation of gravitation should reduce to Poisson's equation in the case of a weak stationary gravitational field. Nevertheless, this requirement has already been compromised in the concordance $\Lambda$CDM cosmology and no longer seems mandatory. It should be noted that the Einstein field equations with a non-zero $\Lambda$ do {\it not } reduce to the Poisson equation \cite{weinberg}. 
While, there is no scope in the standard paradigm to mend this shortcoming, the new theory (being a fundamentally different theory wherein matter does not appear explicitly) should not be compared with the Poisson equation (wherein matter appears explicitly).

Although this entirely new insight about the 
geometry serving as the source of gravitation in the metric theories of gravity, may appear
orthogonal to the usual understanding, it is not only in striking agreement with the theory
and observations, but also provides natural explanations to some unexplained puzzles.
Additionally, it solves the long-standing problems of the standard cosmology: The flatness problem (requiring
the initial density of matter, represented by the
energy-stress tensor, to be extremely fine-tuned to its
critical value) and the cosmological constant problem (whose origin lies in a conflict
between the energy-stress tensor and the vacuum expectation values derived from the quantum
field theory) are averted due to the absence of the energy-stress tensor from the field equations. Horizon problem is solved, as no horizon exists in the resulting cosmological model given by equation (\ref{eq:milne}), and the whole universe is always causally connected.

Equations $R^{\mu\nu}=0$ get strong supports from observations ranging from the solar system to the universe without requiring the usual epicycles of the standard theory, such as inflaton, non-baryonic dark matter and dark energy. 
 Let us recall that the classical tests of GR consider $T^{\mu\nu}=0$ in Einstein's equations, and hence they have been limited to test $R^{\mu\nu}=0$ only (more specifically, they have tested the predictions of the Schwarzschild and Kerr solutions only). Thus the complete Einstein's equations with a non-vanishing $T^{\mu\nu}$ have never been tested directly in any experiment.


\end{document}